\documentclass[aps,prd,preprintnumbers,superscriptaddress,groupedaddress,nofootinbib]{revtex4}
\pdfoutput=1

\usepackage{multirow}
\usepackage{subfigure}
\usepackage{graphicx}
\usepackage{amsfonts}
\usepackage{amsmath,amssymb}
\usepackage{slashed}
\usepackage{feynmp}
\usepackage{verbatim}
\usepackage{caption}
\usepackage{url}
\usepackage{color}
\usepackage{cancel}

\begin{document}
\title{Dark Matter in Leptophilic Higgs Models After the LHC Run-I}
\author{Matthew R.~Buckley}
\affiliation{Department of Physics and Astronomy, Rutgers University, Piscataway, NJ 08854, USA}
\author{David Feld}
\affiliation{Department of Physics and Astronomy, Rutgers University, Piscataway, NJ 08854, USA}

\date{\today}

\begin{abstract}
We examine the leptophilic two Higgs doublet model with fermionic dark matter, considering the range of experimental constraints on the Higgs sector. The measurements of the 125~GeV Higgs from the LHC Run-I allow us to focus on those remaining processes that may play an important role at colliders. We find that the leptophilic model allows for a much lighter Higgs than in other two-Higgs models, although discovery at the LHC will be difficult.  Adding a dark matter sector motivated by supersymmetric extensions of the leptophilic model, we find the existing parameter space can accommodate constraints from direct detection and the invisible widths of the Higgs and $Z$, while also fitting the Galactic Center gamma ray excess reported by analyses of {\it Fermi}-LAT data. We also discuss the status of the fully supersymmetric version of such models, which include four Higgs doublets and a natural dark matter candidate.
\end{abstract}
\maketitle

\section{Introduction \label{sec:intro}}

The minimal particle content necessary to realize the Higgs mechanism in the Standard Model is one scalar $SU(2)_L$ doublet with hypercharge $\pm 1/2$. The discovery of the 125 GeV Higgs by ATLAS and CMS \cite{Aad:2012tfa,Chatrchyan:2012ufa} is an important step on the path to understanding electroweak symmetry breaking (EWSB) and appears consistent with the minimal Higgs sector. However, we should keep in mind that the full scalar sector remains unknown. Therefore, searches for an extended Higgs sector will be an important part of the LHC Run-II. One of the simplest ways to enlarge this sector is to introduce additional scalar doublets, with the minimal extension being a two-Higgs-doublet model (2HDM). While the best known realization of these additional doublet models occurs in supersymmetry, which requires a minimum of two Higgs doublets and is perhaps the most popular candidate for physics beyond the Standard Model, 2HDMs can be found in a variety of other Standard Model extensions.  For example, models of baryogenesis with two Higgses \cite{Turok:1990zg,Joyce:1994zt,Funakubo:1993jg,Davies:1994id, Cline:1995dg,Cline:1996mga,Laine:2000rm,Fromme:2006cm,arXiv:1106.0790,arXiv:1107.3559} can incorporate enough CP-violation to generate a sufficient baryon asymmetry, while the Standard Model alone cannot~\cite{Trodden:1998qg}.

Whereas the Standard Model Higgs has one physical degree of freedom after EWSB, in the form of a charge-parity (CP) even scalar, 2HDMs have five degrees of freedom: two CP-even neutral scalars ($h$ and $H$), one CP-odd neutral pseudoscalar ($a$), and a charged scalar ($H^\pm$). With two doublets the fermions can couple to the scalar sector in a variety of ways. In this paper we examine a 2HDM in which one Higgs doublet has Yukawa couplings with both the up- and down-type quarks, while the other doublet couples to the leptons. These are known as leptophilic two-Higgs doublet models (L2HDM).\footnote{In the literature the leptophilic 2HDM is also commonly referred to as the Type-X 2HDM. In other works it has also been called ``Model I", ``Model IIA", and ``Model IV."} 

Though studies of leptophilic models (including connections to dark matter) have been carried out previously ~\cite{Barger:1989fj, Grossman:1994jb,Akeroyd:1994ga,Akeroyd:1996di, Barnett:1983mm,Barnett:1984zy, Aoki:2009ha}, after discovery of the 125~GeV Higgs and the early measurements of its couplings to gauge bosons and fermions (see Refs.~\cite{Aad:2014eva,Aad:2014eha,Aad:2014xzb,ATLAS:2014aga,Khachatryan:2014jba,CMS:2014jga,ATLAStau,Aad:2015gra} for experimental results and Refs.~\cite{Low:2012rj,Corbett:2012dm,Giardino:2012dp,Ellis:2012hz,Espinosa:2012im,Montull:2012ik,Carmi:2012in,Plehn:2012iz,Carena:2013ooa,Ellis:2014jta,Craig:2015jba,Corbett:2015ksa} for theory analysis) we can paint a much clearer picture of the available parameter space in 2HDMs, leptophilic and otherwise. In particular, the CP-even Higgs boson at 125~GeV appears to be close to the alignment limit, resulting in properties very similar to those found in a minimal single-Higgs sector. As a consequence, processes which would be potentially significant for production of additional physical Higgs scalars at the LHC in a generic 2HDM are now expected to be highly suppressed.

With this in mind, we consider the available parameter space for a L2HDM, reinterpreting Run-I searches and considering the prospects of future searches at the 13 TeV LHC. We further consider the possibility of coupling fermionic dark matter to the Standard Model through the leptophilic Higgs. So-called Higgs Portal couplings to dark matter are theoretically attractive \cite{LopezHonorez:2012kv, Berlin:2015wwa} and link two of the most pressing open questions in theoretical physics. As we shall show, leptophilic Higgs models greatly relax the existing constraints as compared to other 2HDMs, and suggest specific channels for further searches at the LHC. As fermionic dark matter coupled to the Higgs sector is realized in supersymmetric extensions, in addition to the ``simple'' L2HDM, we further consider a supersymmetric extension, which contains minimally four Higgs doublets.

In addition to the experimental constraints on the Higgs sector and new particles coupling to gauge bosons from the LHC and LEP-II, we consider the leptophilic Higgs contribution of the new scalars to the $(g-2)$ value of the muon, on dark matter from direct detection experiments and from indirect detection searches. In particular, we discuss the dark matter interpretation of the unexplained excess of gamma rays coming from the Galactic Center, found in the {\it Fermi}-LAT data \cite{Abdo:2010ex,GeringerSameth:2011iw,Ackermann:2011wa,Geringer-Sameth:2014qqa}. Fermionic dark matter annihilation mediated by scalars is $p$-wave suppressed, {\it i.e.}~the cross section is $v^2$-dependent at leading order. In this case, the velocity of dark matter today is far too small to account for the necessary cross section needed to explain the gamma ray excess. However, annihilation modes through a pseudoscalar proceed via the velocity-independent $s$-wave channel, and so has potential to fit the anomaly (as has been pointed out in the leptophilic context in Ref.~\cite{Marshall:2011mm}). We find parameter space available to account for the observed excess, compatible with the LHC Higgs results, in both the L2HDM and the full supersymmetric four-Higgs model. The solution requires a light pseudoscalar and large $\tan\beta$, which is also the region of interest for solutions to the $(g-2)_\mu$ anomaly.

This paper is organized as follows. Section~\ref{sec:overview} gives an overview of the leptophilic two-Higgs doublet model. The theory along with constraints from the Higgs sector are described in detail. Section~\ref{sec:dm} introduces the dark matter sector of the L2HDM. The ingredients are presented along with notation and couplings. We follow with discussions of implications for direct/indirect detection experiments, as well as for the thermal relic abundance. Section~\ref{sec:susy} discusses the full supersymmetric model and the key differences between it and a generic L2HDM.

\section{The Leptophilic Two-Higgs Doublet Model \label{sec:overview}}

All models with two or more Higgs doublets must assume some symmetry structure in the Higgs-Standard Model fermion couplings in order to prevent large flavor violation at tree-level. The L2HDM ~\cite{Barger:1989fj, Grossman:1994jb,Akeroyd:1994ga,Akeroyd:1996di, Barnett:1983mm,Barnett:1984zy, Aoki:2009ha} extends the Standard Model to contain a Higgs sector with two complex scalars, and assumes an additional ${\mathbb Z}_2$ symmetry which selects one doublet ($\Phi_L$) to have Yukawa couplings with the leptons only, and the other ($\Phi_Q$) to have Yukawa interactions with both up- and down-type quarks. We first review the notation of the L2HDM, and then consider the experimental constraints on this model. 

\subsection{Higgs Sector}
Our two SU(2)$_L$ doublet scalars we assume have equal and opposite hypercharge, $\Phi_L$ being the doublet with $-1/2$ and $\Phi_Q$ with $+1/2$. Allowing for the possibility of charge-neutral vacuum expectation values (vevs) $v_Q$ and $v_L$, we can expand the fields in terms of the neutral CP-even, neutral CP-odd, and charged components:
\begin{equation}
\Phi_Q = \begin{pmatrix}
  h^+_Q  \\
  \tfrac{1}{\sqrt{2}}(v_Q + h_Q + i a_Q)
 \end{pmatrix}, \quad      \Phi_L = \begin{pmatrix}
  \tfrac{1}{\sqrt{2}}(v_L + h_L - i a_L)    \\
    h^-_L
 \end{pmatrix}.
\end{equation}
Assuming both that CP is conserved in the Higgs sector and that CP is not spontaneously broken when the electroweak symmetry is broken, and imposing a softly-broken ${\mathbb Z}_2$ symmetry under which $\Phi_L$ is odd and $\Phi_Q$ is even, the most general scalar potential is:
\begin{eqnarray}
V(\Phi_Q,\Phi_L) & = & - \mu_Q^2 |\Phi_Q|^2 +\mu_L^2 |\Phi_L|^2 - (\mu_{QL}^2 \Phi_Q \Phi_L + {\rm h.c.})   \nonumber \\
& &  + \frac{\lambda_2}{2} |\Phi_L|^4 + \frac{\lambda_2}{2} |\Phi_Q|^4  + \lambda_3|\Phi_Q|^2|\Phi_L|^2 +\lambda_4 |\Phi_Q \Phi_L|^2+ \left\{ \frac{\lambda_5}{2} (\Phi_Q\Phi_L)^2 +\mbox{h.c.} \right\}  
\end{eqnarray}
where all coefficients are real. Experimentally viable EWSB occurs for the region of parameter space where this potential is minimized by
\begin{equation}
\langle \Phi_Q \rangle
= \left( \begin{array}{c} 0 \\ \displaystyle{\tfrac{v_Q}{\sqrt{2}}} \end{array} \right),
\quad
\langle \Phi_L \rangle
= \left( \begin{array}{c} \displaystyle{\tfrac{v_L}{\sqrt{2}}} \\ 0 \end{array} \right).
\end{equation}
Requiring stability of the vacuum allows us to solve for the mass parameters $\mu_Q$ and $\mu_L$.
%
%
We define the angle $\beta$ by the ratio of the vevs $\tan(\beta) \equiv v_Q/v_L$, and the vevs themselves satisfy $v_Q^2 + v_L^2 \equiv v^2 = (246~\mbox{GeV})^2$. We will typically be interested in the parameter space where $\tan\beta > 1$, which will result in the physical Higgs fields having increased couplings to leptons and suppressed couplings to quarks.

After EWSB, three fields are eaten by the $W$ and $Z$ gauge bosons, leaving five physical degrees of freedom: two neutral CP-even states, a neutral CP-odd state, and a charged state. The angle $\alpha$ rotates the CP-even states into the physical mass eigenstates $H$ and $h$, while the angle $\beta$ rotates the CP-odd state and the charged states into the physical states $a$ and $H^{\pm}$:
\begin{eqnarray} 
a &=& -\sin \beta ~ a_L + \cos \beta ~a_Q \nonumber \\
H^\pm &=& -\sin \beta ~h^\pm_L + \cos \beta ~h^\pm_Q  \\
h &=& - \sin \alpha ~h_L + \cos \alpha ~h_Q \nonumber\\
H &=& \cos \alpha ~ h_L + \sin \alpha ~h_Q \nonumber,
\end{eqnarray}

After diagonalizing the mass terms in the potential, we find the physical masses take the form: 
\begin{eqnarray} 
m_{h,H}^2 &=& \frac{\mu_{QL}^2}{s_{2 \beta}} + \frac{1}{2} v^2 \left(  \lambda_1c_\beta^2 + \lambda_2 s_\beta^2 \right) 
 \pm \frac{1}{2} \sqrt{ \left( \frac{2 \mu_{QL}^2}{t_{2 \beta}  } - v^2 \left( \lambda_1c_\beta^2 - \lambda_2 s_\beta^2 \right) \right)^2 -  \left( 2 \mu_{QL}^2 - v^2 \lambda_{345}  s_{2 \beta}\right)^2 }  \nonumber \\ 
m_a^2 &=& \frac{2\mu_{QL}^2}{s_{2 \beta}}  - v^2\lambda_5  \\ 
m_{H^\pm}^2 &=& \frac{2\mu_{QL}^2}{s_{2 \beta}} - \frac{v^2}{2}\left(\lambda_4+\lambda_5 
\right) = m_a^2 + \frac{v^2}{2}(\lambda_4-\lambda_5),\nonumber
\end{eqnarray}
where we employ the notation $s_x = \sin x$, $c_x = \cos x$, and $t_x = \tan x$, and $\lambda_{345} = \lambda_3 + \lambda_4 + \lambda_5$. 
Note that it is possible to choose these four masses independently in a general 2HDM, though the EWSB precision measurements do require a relatively small mass splitting between the $H$ and $H^\pm$ (see {\it e.g.}~Ref.~\cite{Abe:2015oca}). This differs from the minimal supersymmetric Standard Model (MSSM), where the quartic interactions are set by gauge-couplings via $D$-terms, with
\begin{equation}
\lambda_1^{\rm MSSM} = \lambda_2^{\rm MSSM} = \frac{g_1^2+g_2^2}{4}, \quad \lambda_3^{\rm MSSM} =  \frac{g_2^2-g_1^2}{4}, \quad, \lambda_4^{\rm MSSM} = -\frac{g_2^2}{2},\quad \lambda_5^{\rm MSSM} = 0.  \label{eq:susylike} 
\end{equation}
Here $g_1$ and $g_2$ are the hypercharge and $SU(2)_L$ gauge couplings, respectively. A consequence of this is that the masses of the $a$, $H^\pm$ and either the $h$ or the $H$ are close in mass (at tree level). In particular, it is difficult to separate the mass of the pseudoscalar and one of the CP-even scalars in the MSSM at tree-level.

Assigning the left-handed lepton doublet and the $\Phi_L$ to be odd under some ${\mathbb Z}_2$, the Yukawa interactions are constrained to be (suppressing flavor indices)
\begin{eqnarray}
{\cal L}_{\rm Yuk} & \supseteq  -y_e L_L \Phi_L \bar{e}_R - y_u Q_L \Phi_Q \bar{u}_R- y_d Q_L \Phi_Q^\ast \bar{d}_R +\mbox{h.c.} \label{eq:L2HDMyuk}
\end{eqnarray}
After EWSB, the couplings of the physical Higgs fields to the fermions is given by
\begin{eqnarray} 
{\cal L}_{\rm Yuk} & = & - \frac{m_q}{v} \bar{q} \left(\frac{c_\alpha }{s_\beta } h-\frac{s_\alpha}{c_\beta}H+\frac{i\gamma^5}{t_\beta} a  \right)q - \frac{m_\ell}{v} \bar{\ell}\left(\frac{s_\alpha}{c_\beta}h+\frac{c_\alpha}{c_\beta}H+i\gamma^5 t_\beta  a \right)\ell  \label{eq:higgsfermionEWSB} \\
 & & -\frac{\sqrt{2} V_{ud}}{t_\beta v} \bar{d}\left(  m_u P_L - m_d P_R\right)u \,H^- -\frac{ \sqrt{2} t_\beta}{v}\bar{\nu} \left( m_\ell P_L\right) e\, H^+ + \mbox{h.c.}  \nonumber 
\end{eqnarray}
The dependence of the couplings between the Higgs fields and the Standard Model fields relative to the single-Higgs value is summarized in Table~\ref{tab:couplings}. As $\tan\beta$ is increased, the couplings to the leptons for the non-Standard Model-like scalars are increased, while the quarks are decoupled; thus in this limit the additional fields are truly leptophilic.

\begin{table}[t]
\centering
\begin{tabular}{|c|c|c|c|}
\hline
    & $q$                        & $\ell$                         & $Z/W$               \\ \hline
$h$ & $\cos \alpha / \sin \beta$ & $-\sin \alpha / \cos \beta$ & $\sin(\beta-\alpha)$ \\ \hline
$H$ & $\sin \alpha / \sin \beta$ & $\cos \alpha / \cos \beta$  & $\cos(\beta-\alpha)$ \\ \hline
$a$ & $\pm\cot \beta$               & $\tan \beta$                & $0$                  \\ \hline
\end{tabular}
\quad \quad \quad  $\xrightarrow{\text{alignment}}$ \quad \quad \quad
\begin{tabular}{|c|c|c|c|}
\hline
    & $q$                        & $\ell$                         & $Z, W$               \\ \hline
$h$ & $1$ & $1$ & $1$ \\ \hline
$H$ & $-\cot \beta$ & $\tan \beta$  & $0$ \\ \hline
$a$ & $\pm\cot \beta$               & $\tan \beta$                & $0$                  \\ \hline
\end{tabular}
\caption{A summary of couplings between the Higgs scalars and Standard Model fermion and gauge boson pairs. There is an implied $\frac{m_f}{v}$ multiplying the fermionic table entries, while the gauge boson entries should be multiplied by their Standard Model couplings. 
The pseudoscalar $a$ has a $+\cot\beta$ prefactor in the coupling to $u$-type quarks and a $-\cot\beta$ to $d$-type. The right table shows the couplings in the alignment limit $ \cos \left( \beta - \alpha \right) = 0$. \label{tab:couplings}}
\end{table}

In the case of a generic 2HDM, there is no reason {\it a priori}  to prefer that either the heavier $H$ or lighter $h$ is the Standard Model-like Higgs discovered at 125~GeV. However, in the MSSM, the masses of $m_{H^\pm}$, $m_H,$ and $m_a$ are constrained to be of the same order (modulo splittings from EWSB), and thus assigning $H$ as the 125~GeV Higgs would imply relatively light charged Higgses, which would result in large flavor-violating decays in the quark sector \cite{Martin:1997ns}. Thus is it typical to identify the lighter $h$ as the 125~GeV Higgs, and place the $H$, $a$, and $H^\pm$ heavier. As we will show, one of the interesting features of the L2HDM is the relaxation of these bounds, widening the available parameter space to the point where the 125~GeV Higgs can be identified as the heavier CP-even Higgs.

\begin{figure}[t]
\includegraphics[width=.60\columnwidth]{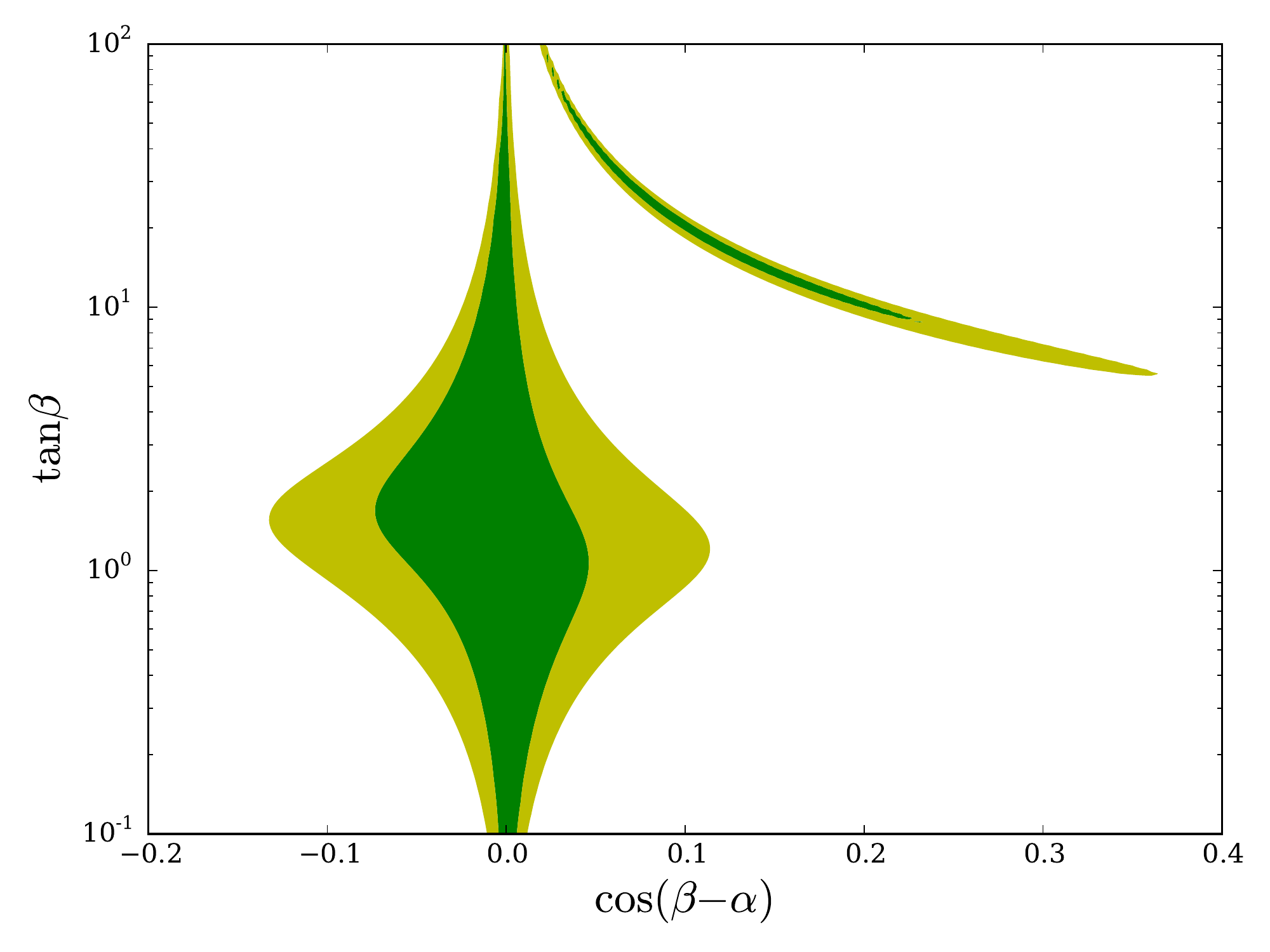}
\caption{$1\sigma$ (green) and $2\sigma$ (yellow) confidence fit of LHC experimental results \cite{Aad:2014eva,Aad:2014eha,Aad:2014xzb,ATLAS:2014aga,Khachatryan:2014jba,CMS:2014jga,ATLAStau,Aad:2015gra} to the L2HDM in the $\cos(\beta-\alpha) - \tan \beta$ plane, assuming tree-level couplings and uncorrelated errors in the experimental measurements. \label{fig:higgsfit}}
\end{figure}

The measurements of the 125~GeV Higgs at the LHC implies that the couplings of the CP-even Higgs to weak bosons must be very close to the single-Higgs Standard Model values. Similarly, couplings to fermions, though less well constrained, are also consistent with the Standard Model prediction, though this is primarily driven by the contribution of the top-Higgs coupling to the loop-induced $h\gamma\gamma$ and $hgg$ widths rather than from direct measurement of associated production. In Figure~\ref{fig:higgsfit}, we show the best-fit contours to the LHC Run-I data as a function of the parameters $\tan\beta$ and $\cos\beta-\alpha$ for the L2HDM, assuming tree-level couplings.\footnote{Fit technique is motivated by Ref.~\cite{Craig:2015jba}.}  For notation simplicity however, we will in general refer to the 125~GeV Higgs as the $h$. The experimental data  indicates that the 125~GeV $h$ is close to the alignment limit, where $\cos(\beta-\alpha) \sim 0$. Note that had we assigned the $H$ as the 125~GeV particle the only replacement in this fit would have been $\cos(\beta-\alpha) \leftrightarrow \sin(\beta-\alpha)$. In this work, we will assume that the $h$ lives exactly in the alignment limit; though minor relaxations of this assumption within the experimental best-fit regions will not significantly change our conclusions.

In the decoupling limit where $m_a, m_{H^\pm}$, and $m_H$ are large $(\gg m_Z)$, the couplings of the lighter Higgs automatically become Standard Model-like. Alignment is possible without decoupling (see {\it e.g.}~Ref.~\cite{Carena:2013ooa}), but one may need to appeal to some mechanism in order to explain the observed similarity of the 125~GeV Higgs to the Standard Model predictions. 

The 125~GeV $h$ can decay into a pair of pseudoscalars, if the $a$ is kinematically accessible ({\it i.e.}~$m_a < 62.5$~GeV). In the alignment limit, the relevant coupling is 
\begin{equation}
g_{haa} = \frac{v}{8}\left(-c_{4\beta}\left(\lambda_1+\lambda_2-2 \lambda_{345} \right)
+\lambda_1+\lambda_2+6\lambda_{345}-16\lambda_5 \right),
\end{equation}
where $\lambda_{345} = \lambda_3+\lambda_4+\lambda_5$. This will result in a $h \to a a$ partial width of 
\begin{equation}
\Gamma(h \to a a)= \frac{g^2_{h a a}}{32 \pi m_h} \left( 1- \frac{4 m^2_\chi}{m^2_h} \right)^{1/2}.
\end{equation}
Given the experimental constraint on the total width of the 125~GeV Higgs $\Gamma_h \lesssim 17$~MeV \cite{CMS:2014ala}, and the branching ratio into non-Standard Model channels must be $\lesssim 50\%$~\cite{CMS:2014ala, ATLAS:2013pma, CMS:2013yda, CMS:1900fga, CMS:2013bfa} this coupling is limited to be $|g_{haa}| \lesssim (11~\mbox{GeV}) \times (1-4m_a^2/m_h^2)^{-1/4}$.  However, if we instead assume that this is the only significant contribution to the Higgs width on top of the Standard Model decays, then new physics contribution to the width can be no larger than $\sim 5$~MeV, and the limit becomes $|g_{haa}| \lesssim(8.0~\mbox{GeV}) \times (1-4m_a^2/m_h^2)^{-1/4}$.   Assuming supersymmetric-like couplings, as in Eq.~\eqref{eq:susylike} results in $g_{haa} = (-34~\mbox{GeV}\cos^22\beta) \times (1-4m_a^2/m_h^2)^{-1/4}$, which is well above the minimum for $\tan\beta$ not near $1$. As a result, without significant loop-corrections to the couplings, supersymmetric-like models are constrained to have $m_a \gtrsim 62.5$~GeV. 

\subsection{Charged Higgs Constraints}

At this time,  the best direct limits on the charged Higgs scalars come from LEP searches \cite{Abbiendi:2003ji}, which place a lower limit of $m_{H^\pm} > 92$ GeV.  The majority of LHC searches for charged Higgses rely on the $H^\pm$ coupling to top quarks \cite{CMSchargedhiggs1,CMSchargedhiggs2, CMSchargedhiggs3, Aad:2014kga}, which is $\tan\beta$ suppressed in the L2HDM.\footnote{The exception is Ref.~\cite{Aad:2015nfa}, which uses $H^\pm-WZ$ couplings, but the cross section to which this search is sensitive is too large to place significant constraints on the 2HDMs considered in this paper.}  Direct charged Higgs pair production followed by decays into $\tau^+\tau^- +\slashed{E}_T$ is searched for, but existing limits are not yet sensitive to new scalars with masses at the LEP-II bound \cite{Aad:2014yka}. As a result, no limits from the LHC can be set on charged Higgs from the L2HDM for $\tan\beta>1$ as of yet.

Indirect limits can be derived from precision measurements of rare decays such as $B \to s \gamma$. In a Type-II 2HDM (as found in the MSSM) where one Higgs doublet couples to up-type quarks and the other to down-type quarks and leptons, the physical charged Higgs couples to up-type quarks as $\cot \beta$ and to the down-type quarks as $\tan \beta$. In contrast, the couplings of the $H^\pm$ to both up and down-type quarks in the L2HDM goes as $\cot \beta$ (see Table~\ref{tab:couplings} and Eq.~\eqref{eq:higgsfermionEWSB}). 

\begin{figure}[t]
\includegraphics[width=.60\columnwidth]{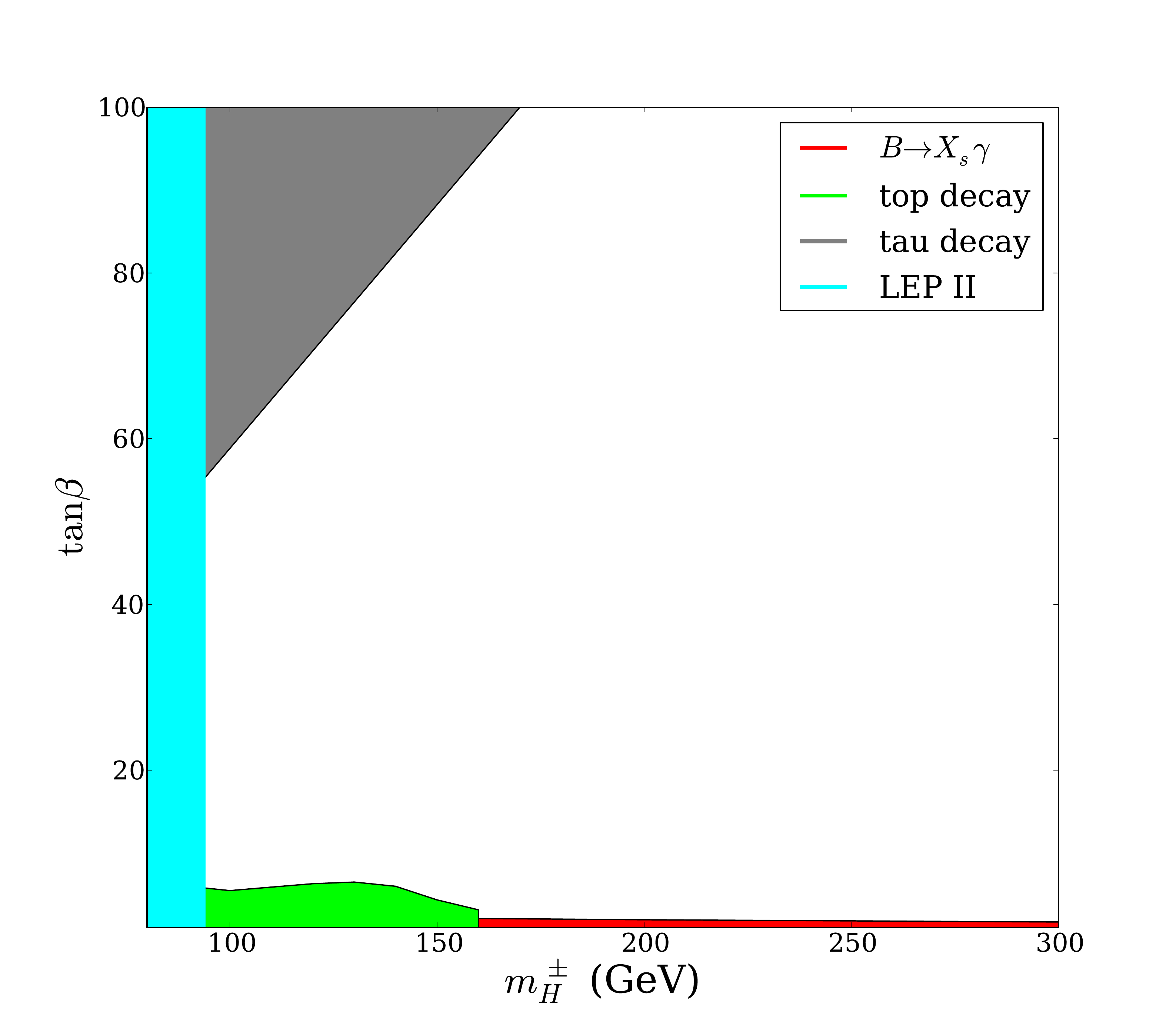}
\caption{Constraints on charged Higgs mass $m_{H^\pm}$ in the LH2DM from LEP II results \cite{Abbiendi:2003ji}, $\tau$ decay \cite{Krawczyk:2004na}, top decay  \cite{Aad:2014kga}, and $B \to X_s \gamma$ (calculated using the procedure described in Ref.~\cite{Ciuchini:1997xe}). \label{fig:chargedhiggs} } 
\end{figure}

Therefore, in the Type II/MSSM case the $H^+ \bar{u} d$ coupling cannot be fully suppressed in any limit of $\tan \beta$, forcing the $H^\pm$ to be heavier than $\sim 300$~GeV for all values of $\tan\beta$ (barring unnatural cancellations in loops between the charged Higgs and new light superpartners) \cite{Haisch:2008ar}. In the L2HDM the coupling to all quarks can be made arbitrarily small as  $\tan \beta$ is increased. For large $\tan\beta$, the loop-induced contributions of $H^\pm$ to tau lepton decay become important \cite{Krawczyk:2004na}, but as shown in Figure~\ref{fig:chargedhiggs}, the available parameter space for charged Higgs in the L2HDM is significant. Figure 2 in Ref. \cite{Haisch:2008ar} shows the analogous plot for the Type II 2HDM. Comparing the two  it is evident that the L2HDM allows for a much larger parameter space than the Type II case.

As the charged Higgs couples to the Standard Model-like Higgs, it will contribute to the loop-induced $h\gamma\gamma$ decay. This coupling is proportional to $\lambda_3$, and so for arbitrary L2HDM models can significantly decrease the diphoton width of the 125~GeV Higgs \cite{Abe:2015oca}. However, for MSSM-like couplings as in Eq.~\eqref{eq:susylike}, this loop contribution is small, resulting in only a $\sim 3\%$ decrease in the diphoton decay width for a charged Higgs with mass at the LEP-II limit. This is in mild tension with the measured value, as the CMS and ATLAS measured $h\to \gamma\gamma$ rate is higher than the Standard Model value \cite{CMS:ril, ATLAS:2013sla}.

\subsection{Muon g-2}

As seen in the limits on the charged Higgs, the presence of light particles in the spectrum can contribute to Standard Model processes at the loop-level. While in almost all cases, the spectacular success of the Standard Model in matching observations means that these loop contributions can only be used to set limits on new physics, the measured value of $(g-2)_\mu$ differs from the Standard Model prediction by over $3 \sigma$: $\Delta a_\mu = (262 \pm 85) \times 10^{-11}$ \cite{Wang:2014sda} where $a_\mu = \left(g-2 \right)_\mu /2 $.  
New pseudoscalars and scalars introduce contributions to $(g-2)_\mu$ at both one- and two-loops, which might help account for this discrepancy. We calculate these corrections as a function of mediator mass and $\tan\beta$ following Eqs.~$(8-11)$ in Ref.~\cite{Wang:2014sda}, for both a CP-even scalar $H$ and CP-odd pseudoscalar $a$ in the L2HDM.

\begin{figure}[h]
\includegraphics[width=.80\columnwidth]{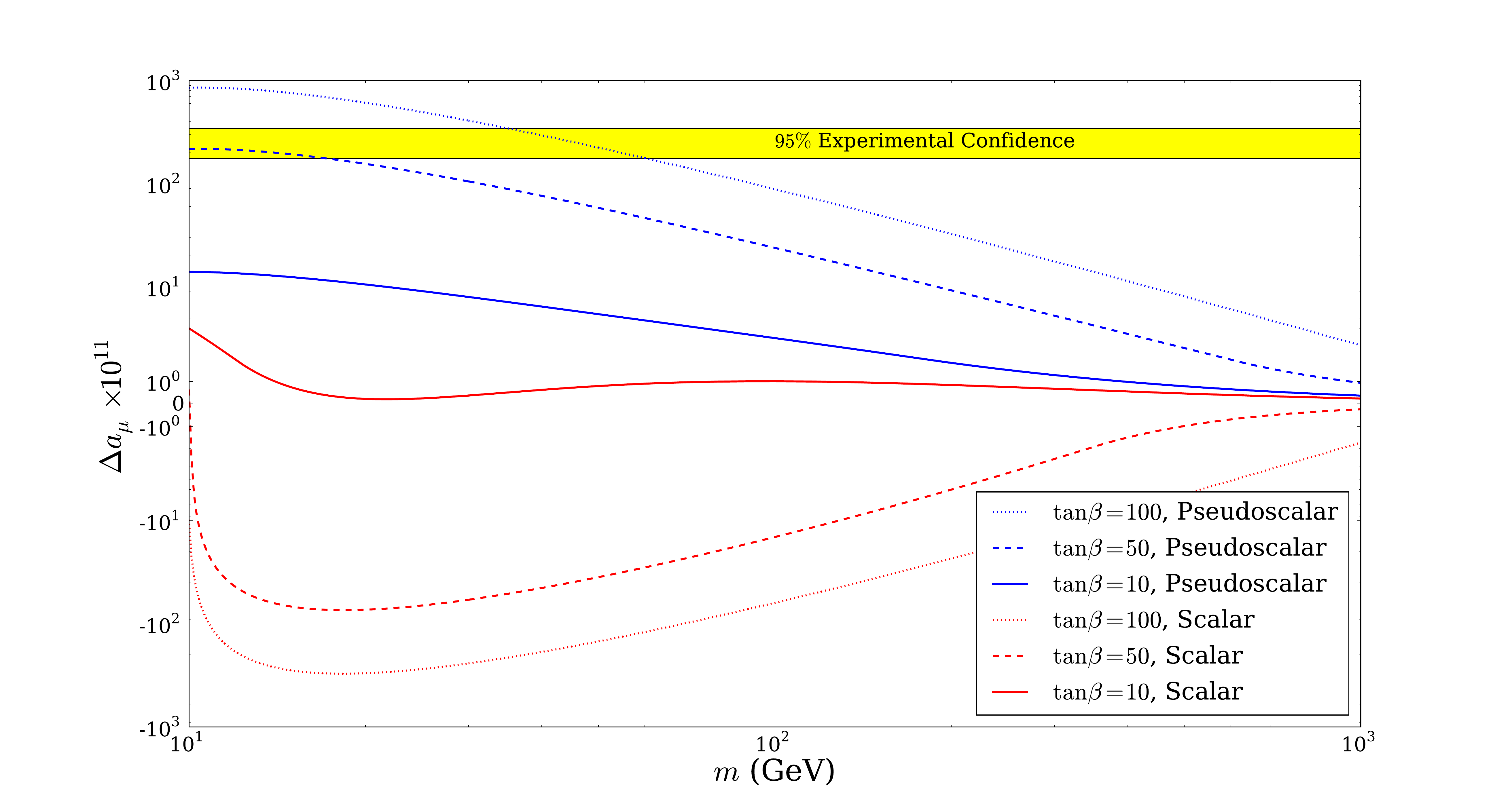}
\caption{Deviation of $a_\mu$ from Standard Model value ($\Delta a_\mu$) due to L2HDM scalars and pseudoscalars as a function of scalar mass for various values of $\tan \beta$. Also shown is the experimental best-fit region  \cite{Wang:2014sda}. \label{fig:g2muon} }
\end{figure}

The loop corrections from the scalar or pseudoscalar are proportional to its coupling with the muon. Since the pseudoscalar $a$ couples to leptons proportional to $ \tan \beta$ in the L2HDM, the contribution from pseudoscalar can greatly enhance  $\left(g-2 \right)_\mu$. In Figure~\ref{fig:g2muon} we plot the contribution of the scalars and pseudoscalars to $a_\mu$ as a function of scalar mass for various values of $\tan \beta$. Pseudoscalars contribute positively to the deviation from the Standard Model value, while scalars contribute negatively for large $\tan\beta$. We see that a light pseudoscalar with high $\tan \beta$ can account for the anomalous measurement. Note that the scalar contribution is negative, however at high masses this contribution is negligible. However, for the L2HDM to be successful in explaining the observed $(g-2)_\mu$ anomaly, a large mass splitting between the light pseudoscalar and the scalars is necessary. Though this is possible in a general L2HDM, it is difficult to accomplish in supersymmetric models without large loop corrections.
Our results agree with those found in the recent work Ref.~\cite{Abe:2015oca}. 

\subsection{Direct Collider Searches}

As with any proposed new particle with electroweak-scale masses, searches for heavy neutral bosons in extended Higgs sectors have been a focus at colliders. Unfortunately, since we now know we live in the limit $\sin( \beta - \alpha) \approx 1$, many processes which would contribute significantly to production of the additional CP-even scalar $H$ in a generic 2HDM
are now known to be highly suppressed. This is because the couplings to gauge bosons of the $H$ is proportional to $\cos(\beta-\alpha)$, so the $W/Z+H$ associated production and vector boson fusion mechanisms are effectively eliminated  in the alignment limit.

The production of the CP-even $H$ in association with the pseudoscalar $a$ through the $ZHa$ coupling is not suppressed in the alignment limit, and is in fact maximal. Searches for neutral Higgses done by LEP-II looked for neutral boson production from the process $e^+ e^- \to Z^* \to H(h)~a$, assuming each decays into pairs of taus (as will be the dominate decay in the L2HDM). This search concluded that there cannot exist both a light pseudoscalar and a light scalar which are kinematically accessible at LEP-II \cite{Schael:2006cr}. Roughly speaking, the sum of $H$ and $a$ masses are required to be greater than $180-200$ GeV. At this time, this LEP-II bound places the only significant limit on the mass of the $H$ and $a$, given that the L2HDM must reside in the alignment limit. As with resolving the $(g-2)_\mu$ anomaly, we find that very light scalars or pseudoscalars are possible, but not both. 

The production of the $H$ or $a$ via gluon-fusion is also suppressed in the L2HDM (assuming $\tan \beta> 1$), as this vertex is induced via top-loops and is therefore proportional to $\cot\beta$. Associated production via top or bottom quark pairs is similarly reduced, as was the case with $H^\pm$ as discussed previously. One might consider associated production of $H$ and/or $a$ alongside tau pairs, but large production cross sections can only be obtained for very light $m_a$ or $m_H$, and for $\tan\beta$ so large that $\tan\beta \times m_\tau/v \gtrsim 1$. This leaves the associated production of $H$ and $a$ through the unsuppressed coupling to $Z$ as the main production mechanism at the LHC. For $\tan\beta \gtrsim 3$, the decay to tau pairs dominates over other fermionic channels for both the $H$ and the $a$, so both of these production mechanisms would result in four tau leptons in the final state. If $m_a < m_H/2$, the channel $H \to a a$ can have a large branching ratio, resulting in six-tau final states. 

The multi-tau final state falls under the multilepton search category at the LHC \cite{Khachatryan:2014jya}. We simulate production of L2HDM scalars and pseudoscalars using \textsc{MadGraph5}~\cite{Alwall:2011uj,Alwall:2014hca} and \textsc{Pythia6}, and \textsc{Delphes3} \cite{deFavereau:2013fsa} for detector simulation. We find that the existing CMS multilepton search performed on 20~fb$^{-1}$ of 8 TeV data does not yet have sufficient sensitivity to place limits on the production of leptophilic Higgs scalar/pseudoscalar pairs with masses at the LEP-II threshold. The 13~TeV Run-II will have signal production cross sections larger by a factor of two, but with similar increases in the primary irreducible dibosons backgrounds. Therefore, discovery of a leptophilic $H$ and $a$ at the LHC would require additional improvements in the multi-tau final states in order to extract the small signal from the background (see {\it e.g.}~Ref.~\cite{Chun:2015hsa}).

\section{Dark Matter \label{sec:dm}}
We now consider the addition of dark matter to the leptophilic models. Both scalar \cite{Logan:2010nw, Boucenna:2011hy} and fermionic \cite{Aoki:2008av} dark matter have been considered in this context previously, but as in the broader consideration of the L2HDM, the discovery and measurement of the 125~GeV Higgs suggests a reassessment of the possibilities with the new evidence in mind. In this work we assume fermionic dark matter, as we are motivated by possible connections of L2HDM models with supersymmetry, where the dark matter would be fermionic neutralinos. In our non-supersymmetric L2HDM, we must add dark matter in by hand, by introducing two $SU(2)_L$ doublets with hypercharge $+1/2$ and $-1/2$, as well as a gauge singlet (these fields are analogous to higgsinos and binos in a supersymmetric model):
\[
\psi_1 = \left(\begin{array}{c} \psi_1^+ \\ \psi_1^0 \end{array}\right),~\psi_2 = \left(\begin{array}{c} \psi_2^0 \\ \psi_2^- \end{array}\right),~\psi_S.
\]
Charging the $\psi_1$ as odd under the same ${\mathbb Z}_2$ as in the Higgs potential, and allowing only soft-breaking terms, the dark matter-Higgs Lagrangian for the neutral states then takes the form
\begin{eqnarray}
{\cal L}_{\psi} & \supseteq   -y_{1} \bar{\psi}_1  \phi_L \psi_S -y_{2} \bar{\psi}_2  \phi_Q \psi_S- \frac{1}{2} m_S \psi_S \psi_S - \mu \psi_1 \psi_2+\mbox{h.c.} 
\end{eqnarray}
The couplings $y_{1}$ and $y_{2}$ are, in a general theory, free parameters. For specificity, and in order to draw a connection with supersymmetric models, we will assume $y_{1} = - y_{2} = g^\prime \sim 0.35$ for numeric calculations, as would be the case for supersymmetric higgsinos and binos coupling to the Higgs sector through the superpotential. A full supersymmetric version would give rise to higgsino and gaugino-matter couplings with Yukawa couplings $\propto g^\prime$. After EWSB, this gives a mass matrix for the neutral components of 
\begin{equation}
\frac{1}{2}\left( \begin{array}{ccc} \psi_1^0 & \psi_2^0 & \psi_S \end{array}\right)\left( \begin{array}{ccc}  0 & -\mu &  \frac{y_{1}v c_\beta}{\sqrt{2}} \\ -\mu & 0 &  \frac{y_{2}v s_\beta}{\sqrt{2}}  \\  \frac{y_{1}v c_\beta}{\sqrt{2}} &  \frac{y_{2}v s_\beta}{\sqrt{2}}  & 2m_S  \end{array} \right)\left( \begin{array}{c} \psi_1^0 \\ \psi_2^0 \\ \psi_S \end{array}\right) + \mbox{h.c.}
\end{equation}
This matrix can be diagonalized by a unitary transformation. We relate the original fields to the mass eigenstates with the coefficient matrix:
\[
\left( \begin{array}{ccc}  N_{11} & N_{12} & N_{13} \\ N_{21} & N_{22} & N_{23} \\ N_{31} & N_{32} & N_{33}  \end{array} \right)\left( \begin{array}{c} \chi_1 \\ \chi_2 \\ \chi_3 \end{array}\right)=\left( \begin{array}{c} \psi_1^0 \\ \psi_2^0 \\ \psi_S \end{array}\right)
\]
where the eigenstate $\chi_1 \equiv \chi$ corresponds to the lightest neutral particle, and thus is a dark matter candidate.
Note that the coefficient $\mu$ will set the mass of the charged states. To satisfy LEP-II constraints on new charged particles, we take $\mu > 100$ GeV for the remainder of the paper.

In Dirac notation, the dark matter $\chi$ has interactions with the physical Higgs fields and the $Z$ boson of
\begin{eqnarray}
{\mathcal L}_{\rm int} &\supseteq&  H \bar{\chi} \chi  \frac{  N_{31}}{\sqrt{2}}\left(  c_\alpha y_{1} N_{11}  +  s_\alpha y_{2} N_{21}  \right) -h \bar{\chi} \chi  \frac{ N_{31}}{\sqrt{2}}\left(  s_\alpha y_{1} N_{11} -  c_\alpha y_{2} N_{21}  \right) \label{eq:DMint} \\
 & & +i a \bar{\chi} \gamma^5 \chi  \frac{N_{31}}{\sqrt{2}}\left( s_\beta  y_{1} N_{11}  + c_\beta  y_{2} N_{21}  \right)- \frac{g}{4c_w} \left[ |N_{11}|^2- |N_{21}|^2  \right] (\bar{\chi} \gamma^\mu \gamma^5 \chi) Z_\mu. \nonumber
\end{eqnarray}
These terms define the dark matter couplings $g_{h\chi\chi}$, $g_{H\chi\chi}$, $g_{a\chi\chi}$ and $g_{Z\chi\chi}$. Note that in the alignment and large $\tan\beta$ limit, both the $H$ and $a$ couple primarily to the $N_{11}$ component (the ``leptophilic higgsino,'' if viewed as a supersymmetric model), while the $h$ couples mostly to the $N_{12}$ component (the ``quark-like higgsino''). 

As the dark matter arises from mixing $SU(2)_L$ doublets with singlets, it couples to both the 125~GeV Higgs and the $Z$, and so can contribute to the invisible width of these particles. The direct decays have widths
\begin{eqnarray}
\Gamma(h \to \chi \bar{\chi}) &= &\frac{g^2_{h \chi \chi} m_h}{8 \pi} \left( 1- \frac{4 m^2_\chi}{m^2_h} \right)^{3/2}, \\
\Gamma(Z \to \chi \bar{\chi}) &=& \frac{g^2_{Z \chi \chi} m_Z}{24 \pi} \left( 1- \frac{4 m^2_\chi}{m^2_Z} \right)^{3/2}.
\end{eqnarray}

\begin{figure}[h]
\includegraphics[width=.60\columnwidth]{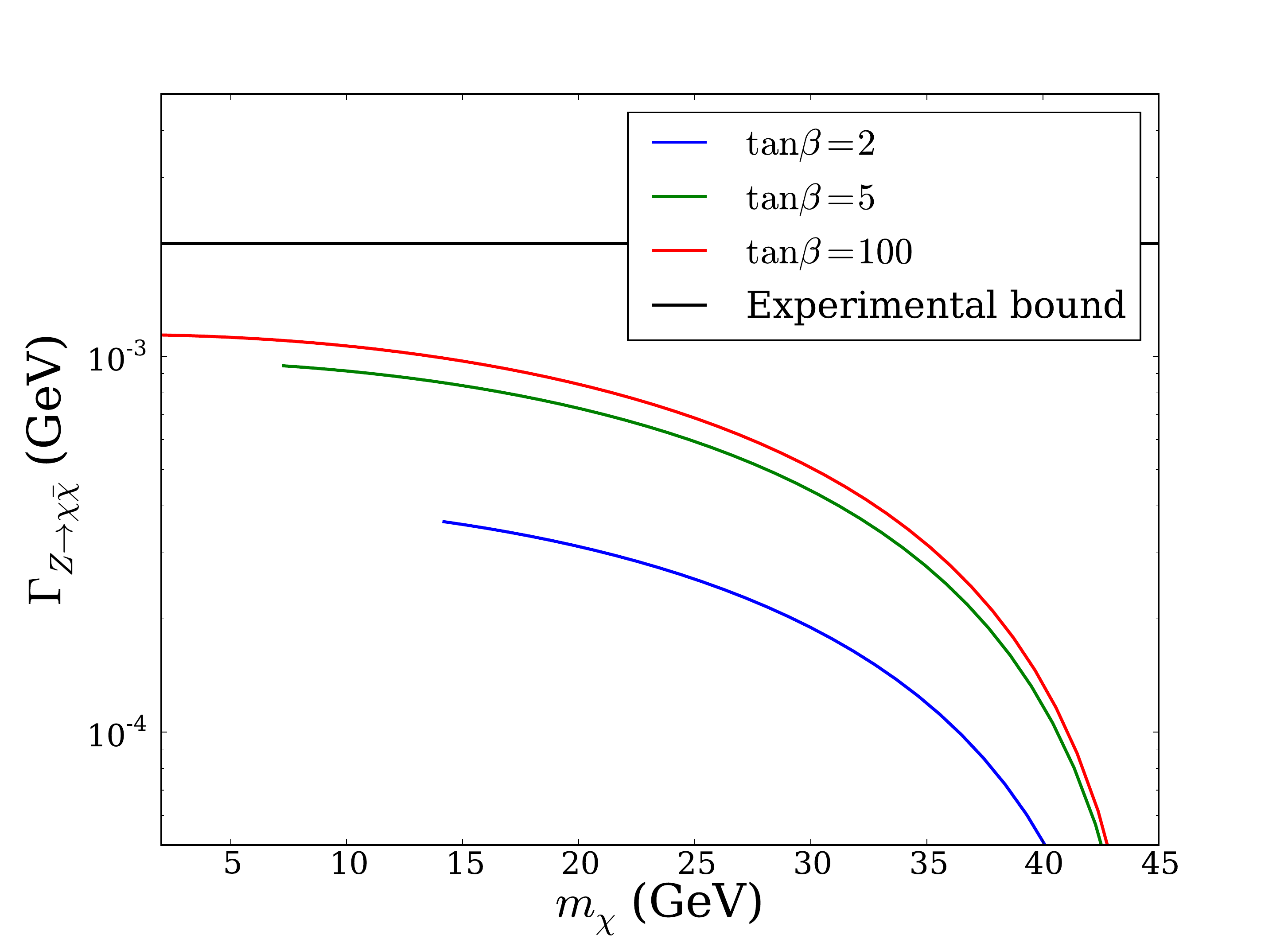}
\caption{Invisible Z width as a function of dark matter mass for various values of $\tan \beta$ with $\mu = 200$ GeV. The limit is $\Gamma_{Z \rightarrow \bar{\chi} \chi} \lesssim 2$ MeV \cite{ALEPH:2005ab}. \label{fig:Zwid}}
\end{figure}

\begin{figure}[h]
\includegraphics[width=.63\columnwidth]{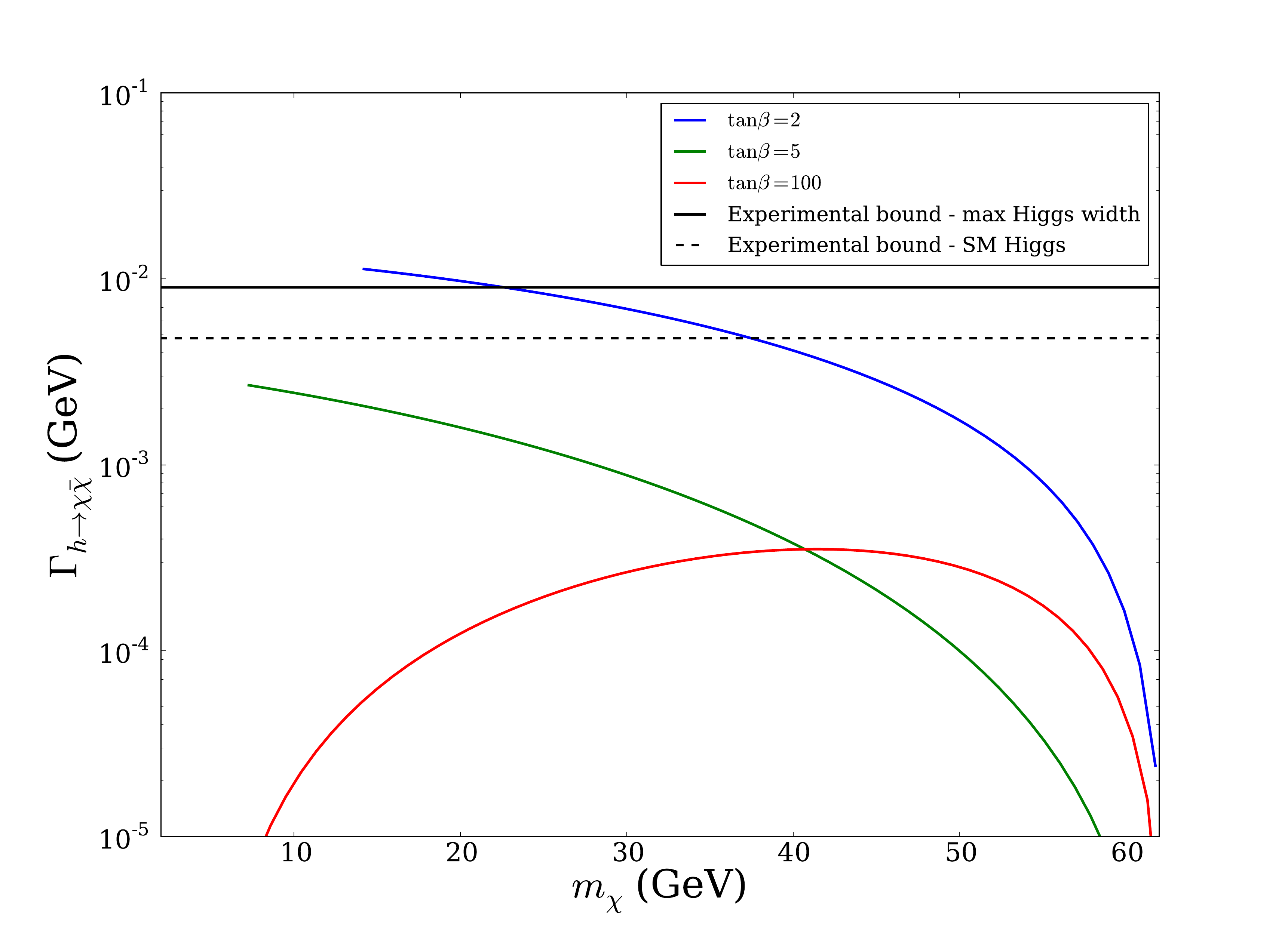}
\caption{Invisible Higgs width as a function of dark matter mass or various values of $\tan \beta$ with $\mu = 200$ GeV. The limit derived from the maximum Higgs width is $\Gamma_{h \rightarrow \bar{\chi} \chi} \lesssim 9$ MeV \cite{CMS:2014ala}, while the limit derived from the SM Higgs width is $\Gamma_{h \rightarrow \bar{\chi} \chi} \lesssim 4.8$ MeV \cite{CMS:2014ala, ATLAS:2013pma, CMS:2013yda, CMS:1900fga, CMS:2013bfa}. \label{fig:higgswid}}
\end{figure}

As discussed previously, the upper limit of the invisible branching ratio is $54\%$ of the total Higgs width \cite{CMS:2014ala, ATLAS:2013pma, CMS:2013yda, CMS:1900fga, CMS:2013bfa}. The upper limit on the total Higgs width is $\sim 17$ MeV \cite{CMS:2014ala}, and so at most we must require $\Gamma_{h_{\rm inv}} \lesssim 9$ MeV. Assuming that, in addition to the dark matter interaction, all the $h$ couplings  are exactly those of the Standard Model, then the limit on the invisible branching ratio translates to $\Gamma_{h_{\rm inv}} \lesssim 5$ MeV. The uncertainty in the $Z$ width constrains $\Gamma_{Z \rightarrow \bar{\chi} \chi} \lesssim 2$ MeV \cite{ALEPH:2005ab}. The invisible widths for $Z$ and $h$ as a function of dark matter mass are shown in Figures~\ref{fig:Zwid} and \ref{fig:higgswid} respectively, assuming $\mu = 200$~GeV. As $m_\chi$ increases, the $SU(2)_L$ component of the dark matter decreases, and so the constraints are more easily satisfied. Note that $m_\chi$ cannot be made arbitrarily small for fixed values of the Yukawas $y_1,y_2$ and $\mu$, thus explaining why some lines in Figures~\ref{fig:Zwid} and \ref{fig:higgswid} are cut off at the low mass end. For our choices of parameters shown in these figures, the $g_{h \chi \chi}$ coupling will receive a cancellation from the $Y = + 1/2$ and $Y = - 1/2$ components of the dark matter. This cancellation occurs at lower $m_\chi$ for higher $\tan \beta$, explaining the cancellation in the $\tan\beta = 100$ curve in Figure~\ref{fig:higgswid}.

\begin{figure}[h]
\includegraphics[width=.60\columnwidth]{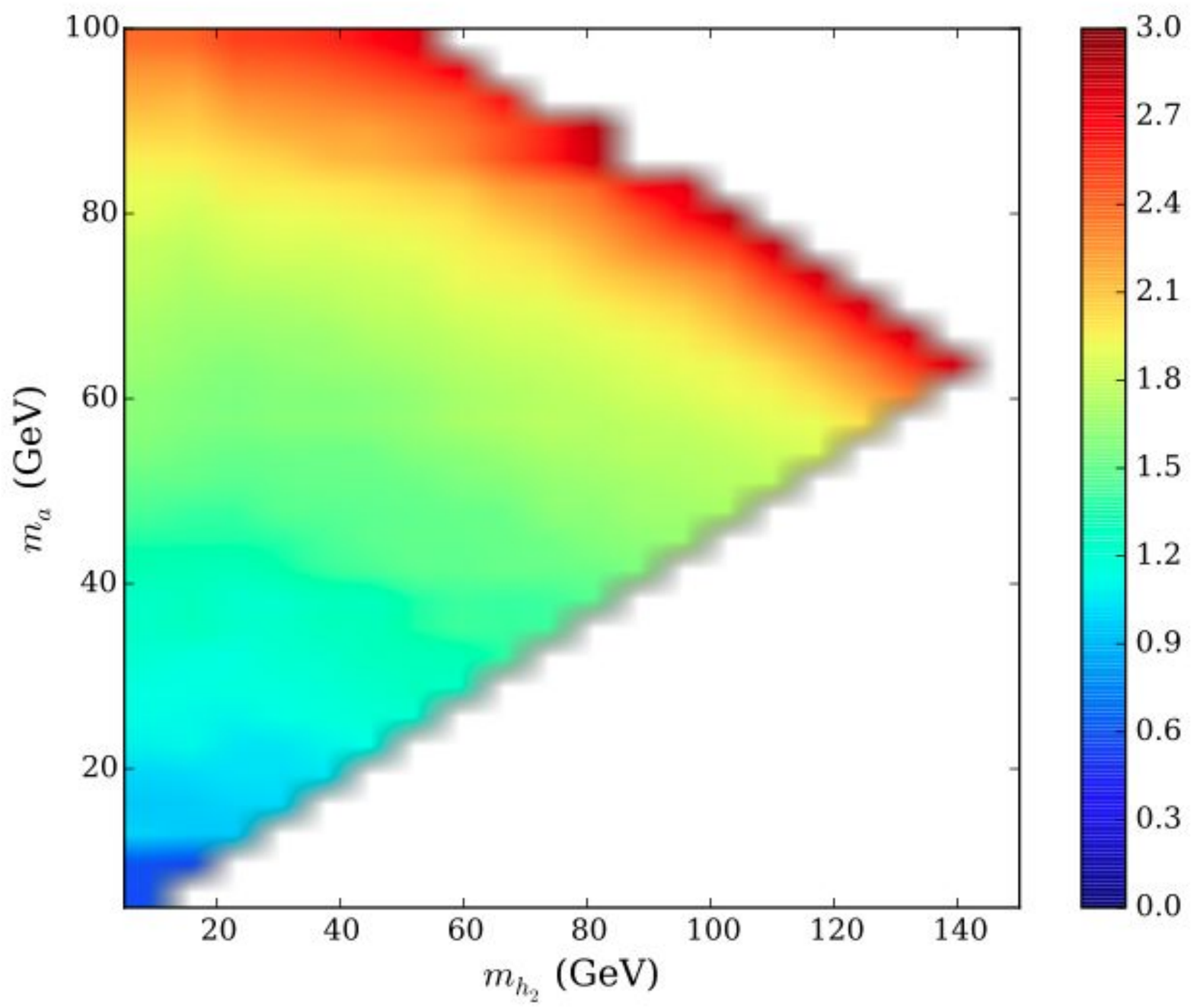}
\caption{Upper limits on $\tan \beta$ from LEP-II constraint \cite{Schael:2006cr} assuming decay into either dark matter or $\tau \tau$, setting the mass of dark matter to 1~GeV and the doublet component $N_{11} = 0.2$. \label{fig:LEPdm}}
\end{figure}

In addition to direct decays of $h$ to dark matter, $h$ could decay to a pair of pseudoscalars, which themselves can decay to dark matter pairs (assuming that each decay is kinematically allowed). The constraints on the $h\to aa$ discussed in the previous section apply here. One might also consider whether invisible decays of the $a$ and $H$ to dark matter could allow the LEP-II bounds on the production $e^-e^+ \to Z^* \to Ha$ \cite{Schael:2006cr} to be evaded, by turning a four-tau final state into one of missing energy only. However, this is possible only for very low values of $\tan\beta$, as the branching ratio of $a$ and $H$ to tau leptons increases at large $\tan\beta$, while the coupling to dark matter asymptotes to a constant in this limit. In Figure~\ref{fig:LEPdm}, we show the upper limit on $\tan\beta$ from LEP-II as a function of scalar and pseudoscalar mass, for 1~GeV dark matter with the maximal doublet component which is in agreement with the invisible width constraints. This choice of parameters maximizes the branching ratio of the $a$ and $H$ into dark matter and so most efficiently avoid the constraints from the tau search at LEP-II.

\subsection{Direct Detection}

Direct detection experiments measure the recoil energy from WIMP-nucleus scattering, placing an upper limit on the dark matter-nucleon elastic scattering cross section. This allows us to place bounds on the parameters that make up the dark matter-scalar and quark-scalar couplings. Dark matter-nucleus scattering is dominated by mediation from the CP even scalars. The pseudoscalar mediated interaction is velocity suppressed and so has negligible contribution to the scattering cross section. Dark matter-nuclei interactions mediated by scalars induce a spin-independent cross section which is constrained by a number of experiments. Presently, the strongest bounds come from the LUX experiment \cite{Akerib:2013tjd} for $m_\chi \gtrsim 6$~GeV and by CDMS-lite  at lower dark matter masses \cite{Agnese:2013jaa}.

The fundamental Lagrangian parameters are translated into dark matter-nucleon scattering cross sections using \cite{Buckley:2013jwa}
\begin{eqnarray}
\sigma_{\chi-p,n} & = & \frac{\mu^2}{\pi} f_{p,n}^2, \\
f_{p,n} & = & \sum_{q=u,d,s} f_q^{p,n}\frac{m_{p,n}}{m_q} \xi_q + \frac{2}{27} f_\text{TG}^{p,n} \sum_{q=c,b,t} \frac{m_{p,n}}{m_q} \xi_q \\
\xi_q &= &\frac{1}{M_H^2}  g_{H q q} ~g_{H\chi \chi} + \frac{1}{M_h^2} g_{h q q} ~ g_{h\chi \chi } ,
\end{eqnarray} 
where $\mu$ is the dark matter-nucleon reduced mass, $\xi_q$ is the effective dark matter-quark coupling, and the parameters $f^{p,n}_q$ and $f^{p,n}_\text{TG}$ are proportional to the quark expectation operators in the nucleon. These must be extracted from lattice QCD simulations \cite{Belanger:2008sj,Young:2009zb,Toussaint:2009pz,Giedt:2009mr,Fitzpatrick:2010em}, and we adopt the values from Ref.~\cite{Fitzpatrick:2010em}. For the purposes of this paper there is no significant difference between the proton and neutron $f_{p,n}$, and so our dark matter scattering is essentially isospin-conserving. 

\begin{figure}[h]
\includegraphics[width=.60\columnwidth]{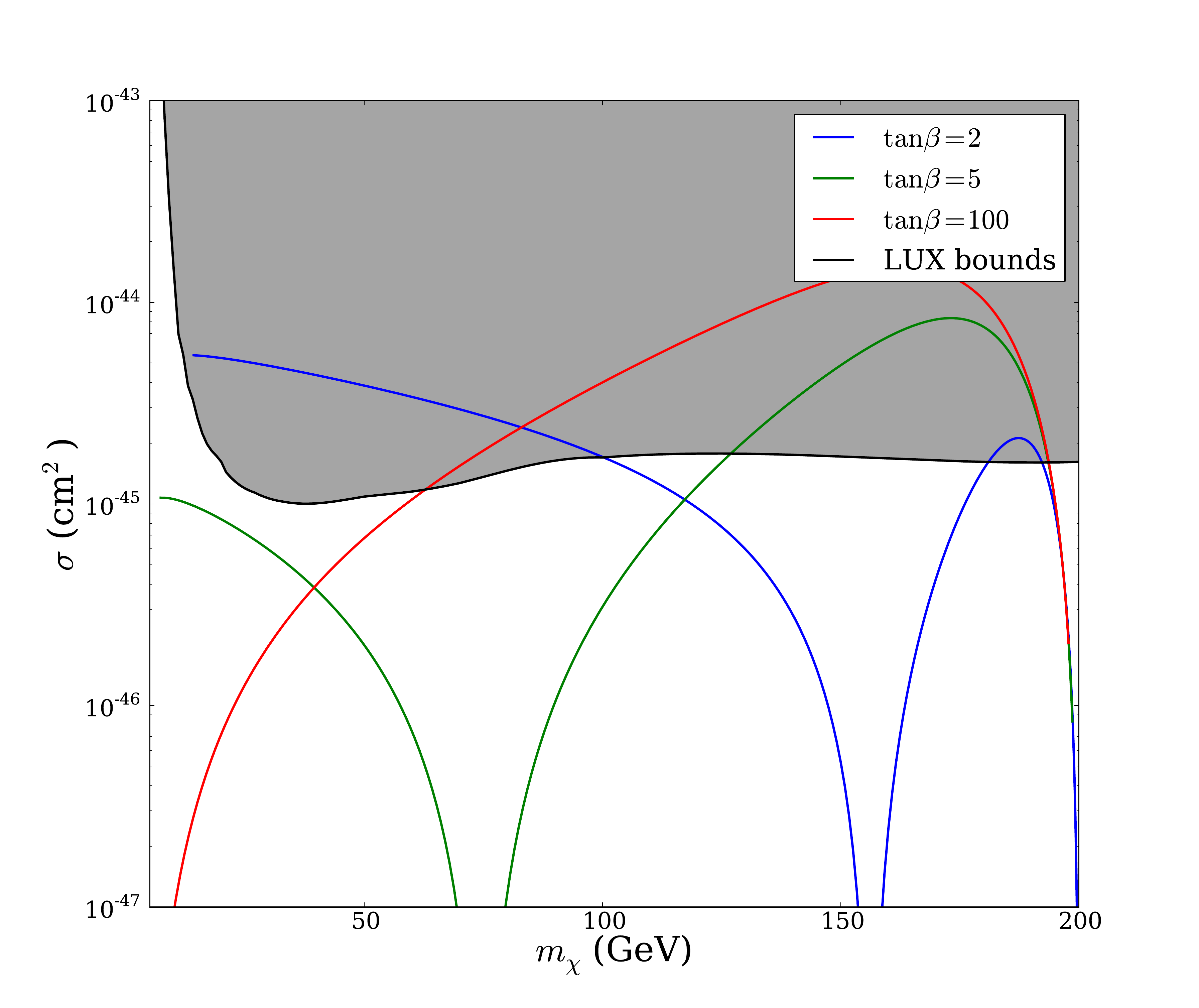}
\caption{Spin-independent cross section as a function of dark matter mass for various values of $\mu$ and $\tan \beta$. Limits from LUX are given in green. We take $m_H = 500$ GeV. \label{fig:dirdet}} 
\end{figure}

In Figure~\ref{fig:dirdet} we show the direct detection cross for various $\mu$ and $\tan \beta$ values as a function of dark matter mass, with $m_H = 500$~GeV (though note that, due to the leptophilic nature of this scalar, a lower mass for $H$ does not significantly change the result). We see that there is space to accommodate LUX bounds, especially as $\mu$ increases. Note again two accidental cancellations which can reduce the measured cross section. One decreases of the $SU(2)_L$ doublet components' contributions to the cross section, which shifts to lower dark matter masses as $\tan \beta$ is increased. The second occurs as $m_\chi \to \mu$ resulting in $N_{13} \to 0$, which we have chosen to be 200~GeV here.

\subsection{Indirect Detection}

The L2HDM with fermionic dark matter model allows for the annihilation of dark matter into Standard Model particles in the Universe today. Indirect detection experiments search for unexplained excesses of gamma rays, positrons, or other Standard Model final states coming from areas of high dark matter density. Whereas direct detection experiments are sensitive to $s$-wave scattering mediated by scalars, thermally averaged annihilation cross sections mediated by scalars are $p$-wave suppressed, {\it i.e.}~proportional to $v^2$. The velocity of dark matter today is very small $( < 10^{-2} c)$, and so $p$-wave annihilation is not detectable. However, annihilation through a pseudoscalar has a velocity independent component and therefore is the main contribution to $ \langle \sigma v \rangle$. The thermally averaged cross section for $(\chi\bar{\chi} \to a^* \to f\bar{f})$  takes the form:
\begin{eqnarray}
\langle\sigma v\rangle & = & \sum_{f} N_f \frac{g_{a \chi \chi}^2 g_{a f f}^2 m_f^2}{2 \pi v^2 \left[ (m_a^2 - 4m_\chi^2)^2 + m_a^2 \Gamma_a^2 \right]} \left [m_\chi^2 \sqrt{1-\frac{m_f^2}{m_\chi^2}} +\frac{3m_f^2}{4 m_\chi \sqrt{1-\frac{m_f^2}{m_\chi^2}} } T \right], 
\end{eqnarray}
where $N_f$ is the number of colors of the fermion $f$, and $T$ is the temperature of the dark matter. We can safely take $T = 0$ since the temperature today is very small. As the $a$ couples primarily to tau leptons in the large $\tan\beta$ limit, which decay into final states with large number of photons, the primary indirect detection search channel will be gamma rays.

Areas of expected high density dark matter in our Galaxy can be found in nearby satellite galaxies \cite{GeringerSameth:2011iw,Ackermann:2012nb,Ackermann:2013yva,Buckley:2015doa}, extra-Galactic sources \cite{Abdo:2010dk,Dugger:2010ys,Ackermann:2012uf,DiMauro:2015tfa,Ackermann:2015tah}, or the Milky Way Galactic Center. A notable excess of gamma rays coming from the Galactic Center has been reported by various analyses of Fermi Gamma-Ray Space Telescope (FGST) data \cite{Abdo:2010ex,GeringerSameth:2011iw,Ackermann:2011wa,Geringer-Sameth:2014qqa}. The source of these gamma-rays is uncertain but can be interpreted as coming from dark matter annihilation. 
Annihilation to taus is one of the final states which fit the excess spectrum reasonably well \cite{Goodenough:2009gk,Hooper:2010mq,Hooper:2011ti,Boyarsky:2010dr,Abazajian:2012pn,Hooper:2012sr,Hooper:2013rwa,Gordon:2013vta,Huang:2013pda,Abazajian:2014fta,Daylan:2014rsa, Calore:2014xka}, when the dark matter mass is in the range $\sim 9-11$ GeV. 
In Figure~\ref{fig:inddet} we show the thermally cross sections as a function of $\tan \beta$ for various values of $\mu$ and $m_a$, fixing $m_\chi = 10$~GeV. The shaded regions are possible values for $\langle \sigma v \rangle$ in our model. As expected the cross section decreases as the mass of the pseudoscalar increases, however we can see that there is still a large range of parameters which fit the Galactic Center excess, even when the pseudoscalar and dark matter masses are far from being in resonance. It is notable that a leptophilic Higgs model furnishes a viable mediator between the visible sector and the dark matter which provides a sufficiently large annihilation cross section; in Type-II Higgs models, a Higgs mediator which can explain the Galactic Center anomaly without a resonance enhancement tends to require additional light states which are well-constrained from direct detection and collider searches \cite{Cheung:2014lqa,Cahill-Rowley:2014ora,Guo:2014gra,Cao:2014efa,Agrawal:2014oha,Caron:2015wda,Butter:2015fqa}. As the mediators considered here are leptophilic, such constraints are easily avoided.

\subsection{Thermal Relic Abundance}

Measurements of CMB anisotropies from telescopes such as Planck give the dark matter component of the Universe's energy density to be $\Omega_\chi h^2 = 0.1187 \pm 0.0017$ \cite{Ade:2013zuv}. The standard Boltzmann relic density calculations imply the corresponding thermal annihilation cross section for dark matter of $\langle \sigma v \rangle \sim 3 \times 10^{-26}$ cm$^3$/s. If we assume that the dark matter is a thermal relic we can calculate the necessary couplings to give the correct thermal annihilation cross section. The calculation is the same as in the case of indirect detection except that we evaluate $T$ at the freeze-out temperature $T_f = m_\chi / 25$ \cite{Buckley:2013jwa}. 

\begin{figure}[t]
\includegraphics[width=.80\columnwidth]{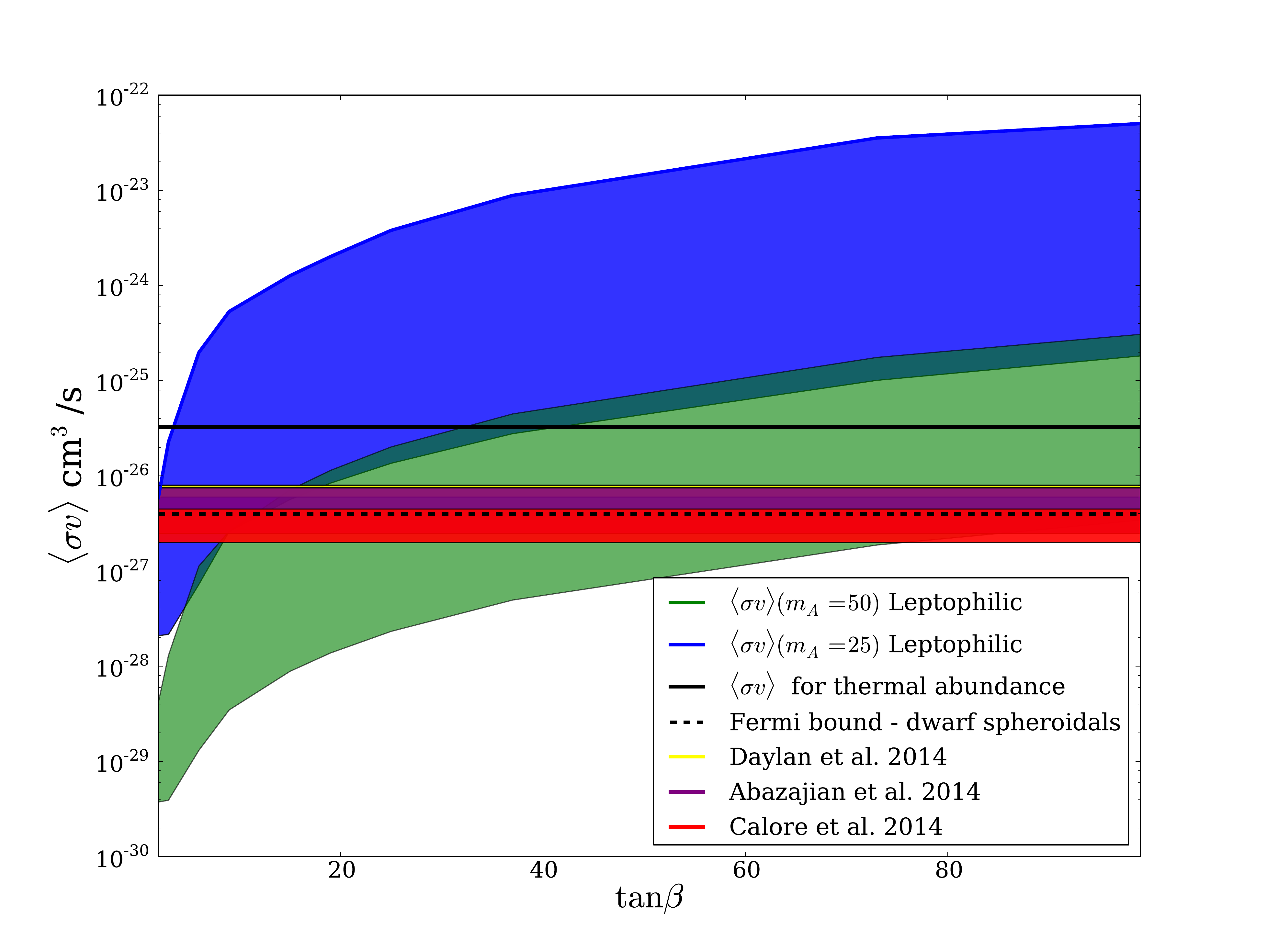}
\caption{Thermally averaged annihilation cross section as a function of $\tan \beta$ for $m_a = 25$ GeV (blue) and $m_a = 50$ GeV (green), assuming $m_\chi = 10$ GeV. Also shown are best fits to the Galactic Center excess in the $\tau \tau$ final state, from Refs.~\cite{Abazajian:2014fta,Daylan:2014rsa, Calore:2014xka}.  \label{fig:inddet}}
\end{figure}

Along with showing available parameter space for indirect detection, Figure~\ref{fig:inddet} also demonstrates the region of parameter space with the correct thermal abundance obtained through couplings with the pseudoscalar. We see that the canonical value of the thermal annihilation cross section can be achieved for many values of $\tan \beta$ and $m_a$, although high values of $m_a$ require increasingly higher values of $\tan \beta$ as expected.

The best fits for the Galactic Center anomaly do not coincide with the thermal relic cross section, a concern that our model shares with other explanations of the excess assuming annihilation to taus. It is possible that additional $p$-wave processes contributed to the thermal relic cross section which are simply inactive today due to velocity suppression. Alternatively, processes going to final states which do not contribute significant gamma-ray flux in the Galactic Center (for example, annihilation to neutrinos) can serve to boost the thermal relic cross section for freeze-out in the early Universe. Our results shown in Figure~\ref{fig:inddet} should be taken to demonstrate that thermal cross sections of the correct magnitude for both thermal relics and the Galactic Center can be relatively easily obtained in the L2HDM with fermionic dark matter, without significant fine-tuning.  

\section{Leptophilic Supersymmetry \label{sec:susy}}

Throughout this paper we have often set unknown couplings in the Higgs and dark matter sectors to their ``supersymmetric'' equivalent in order to both reduce parameter space and to make contact with the most familiar interpretations of two-Higgs doublet models, as realized in the MSSM. However, the L2HDM cannot be supersymmetrized without adding new Higgs fields, as the Yukawa interactions for the Standard Model fermions must come from a holomorphic superpotential, forbidding down quark masses from the $\Phi^*_Q$ term in Eq.~\eqref{eq:L2HDMyuk}. The minimum number of Higgs doublets required for a leptophilic supersymmetric model which contains the Standard Model is four, $H_u$, $H_d$, $H_\ell$, and $H_n$, with the down- and lepton-Higgses having hypercharge $-1/2$ and the up-type and $H_n$ having $+1/2$. This model was introduced in Ref.~\cite{Marshall:2010qi}, and we adopt much of their notation.

Of the four Higgs doublets, we assume as in the L2HDM that the $H_\ell$ and $H_n$ are odd under some ${\mathbb Z}_2$, along with the lepton doublets. Then the superpotential for the theory is (again suppressing flavor indices)
\begin{equation}
W = y_u  Q_L  \bar{u}_R H_u+y_d Q_L \bar{d}_R H_d +y_\ell L_L \bar{ e}_R H_\ell + \tilde{\mu}_1 H_u H_d + \tilde{\mu}_2 H_\ell H_n
\end{equation}
The soft supersymmetric-breaking potential will break the ${\mathbb Z}_2$ symmetry,
\begin{equation}
V_{\rm soft} = \mu^2_u |H_u|^2 + \mu^2_d |H_d|^2 + \mu_\ell^2 |H_\ell|^2+\mu_n^2 |H_n|^2+\left(\mu_1^2 H_u H_d + \mu_2^2 H_n H_\ell + \mu_3^2 H_u H_\ell +\mu_4^2 H_n H_d + \mbox{h.c.} \right).
\end{equation}

The quartic couplings are provided by supersymmetric $D$-terms, as in the MSSM. We assume that each Higgs obtains a CP- and $U(1)_{\rm em}$-conserving vev, with $v^2 = v_u^2+v_d^2+v_n^2+v_\ell^2$. In analogy with the ratio $\tan \beta$, we define three angles: 
\begin{equation}
\tan^2 \beta = \frac{v_u^2+v_d^2}{v_\ell^2+v_n^2}, \quad \tan \gamma_q = \frac{v_u}{v_d}, \quad \mbox{and}\quad \tan \gamma_\ell = \frac{v_n}{v_\ell}.
\end{equation}
As in the L2HDM, the angle $\beta$ describes the amount of the EWSB vev which lies in the lepton sector, and we are primarily interested in the large $\tan\beta$ limit. We refer readers to Ref.~\cite{Marshall:2010qi} for the mass matrices of the Higgs sector after EWSB.

As in the MSSM, the four-Higgs supersymmetric model has a natural dark matter candidate, the lightest neutralino, here made up of winos $\tilde{W}$, binos $\tilde{B}$, and neutral higgsinos: $\tilde{h}_u$, $\tilde{h}_d$, $\tilde{h}_n$ and $\tilde{h}_\ell$. The mass matrix for the dark matter (in the $\tilde{h}_u,\tilde{h}_d,\tilde{h}_n,\tilde{h}_\ell,\tilde{W},\tilde{B}$ basis) is
\begin{equation}
{\cal M}_\chi = \left(\begin{array}{cccccc} 0 & \tilde{\mu}_1 & 0 & 0 & -\frac{g v s_\beta s_{\gamma_q}}{\sqrt{2}} & \frac{g' v s_\beta s_{\gamma_q}}{\sqrt{2}} \\  \tilde{\mu}_1 &  0 & 0 & 0 & \frac{gv s_\beta c_{\gamma_q}}{\sqrt{2}} & -\frac{g'v s_\beta c_{\gamma_q}}{\sqrt{2}} \\  0 & 0 & 0 & \tilde{\mu}_2 & -\frac{gv c_\beta s_{\gamma_\ell}}{\sqrt{2}} & \frac{g'v c_\beta s_{\gamma_\ell}}{\sqrt{2}} \\  0 & 0 &  \tilde{\mu}_2 & 0 & \frac{gv c_\beta c_{\gamma_\ell}}{\sqrt{2}} & -\frac{g'v c_\beta c_{\gamma_\ell}}{\sqrt{2}} \\ -\frac{gv s_\beta s_{\gamma_q}}{\sqrt{2}} & \frac{gv s_\beta c_{\gamma_q}}{\sqrt{2}} & -\frac{gv c_\beta s_{\gamma_\ell}}{\sqrt{2}} & \frac{gv c_\beta c_{\gamma_\ell}}{\sqrt{2}} & M_2 & 0 \\   \frac{g'v s_\beta s_{\gamma_q}}{\sqrt{2}} & -\frac{g'v s_\beta c_{\gamma_q}}{\sqrt{2}} &\frac{g'v c_\beta s_{\gamma_\ell}}{\sqrt{2}} & -\frac{g'v c_\beta c_{\gamma_\ell}}{\sqrt{2}} & 0 & M_1 \end{array} \right)
\end{equation}
This matrix is diagonalized by the $6\times 6$ matrix $N$. Assuming $M_1 \ll \tilde{\mu}_1,\tilde{\mu}_2 \ll M_2$, the lightest dark matter candidate will be primarily $\tilde{B}$ with an admixture of higgsino. The couplings of the neutralinos to the $Z$ are:
\begin{equation}
\frac{g}{c_w} \left(|N_{i,u}|^2+|N_{i,n}|^2-|N_{i,d}|^2-|N_{i,\ell}|^2 \right) (\bar{\tilde{\chi}}_i \gamma^\mu \gamma^5 \tilde{\chi}_i)Z_\mu.
\end{equation}
The couplings to the $i^{\rm th}$ CP-even and CP-odd Higgses are:
\begin{eqnarray}
{\cal L}_{\rm int} & = & -\frac{1}{2} \left[gN_{k,W}-g'N_{k,B}\right] \left[ (U_{i,u} N_{k,u}-U_{i,d} N_{k,d}+U_{i,n} N_{k,n}- U_{i,\ell} N_{i,\ell}) \right] h_i \bar{\tilde{\chi}}_k \tilde{\chi}_k \\ 
& & +\frac{i}{2} \left[gN_{k,W}-g'N_{k,B}\right] \left[ (V_{i,u} N_{k,u}-V_{i,d} N_{k,d}+V_{i,n} N_{k,n}- V_{i,\ell} N_{i,\ell})\right] h_i \bar{\tilde{\chi}}_k \tilde{\chi}_k, \nonumber
\end{eqnarray}
where $U$ and $V$ are the rotation matrices for the CP-even and CP-odd Higgses $h_i$ and $a_i$. 

After EWSB, the physical scalar sector has four CP-even scalars $h_i$, three CP-odd pseudoscalars $a_i$, and three charged scalars $H^\pm_i$. As the Higgs quartics are set by the gauge-interactions of $SU(2)_L$ doublets with hypercharge $\pm1/2$, the CP-even Higgs scalar that has Standard Model-like couplings to the $W$ and $Z$ bosons has a tree-level mass bounded from above by $m_Z$. Therefore, large loop-contributions to the mass of the Standard Model-like Higgs are still required, as in the MSSM, and the Little Hierarchy Problem \cite{Martin:1997ns} remains.  

However, as in the L2HDM, this four-Higgs model does allow for the possibility that the 125~GeV Higgs is not the lightest Higgs scalar, unlike in the MSSM. As discussed in Section~\ref{sec:overview}B, in a Type-II 2HDM, as in the MSSM, the charged Higgs cannot be lighter than $\sim 300$~GeV without requiring light superpartners who's contribution to $b\to s\gamma$ is fine-tuned to cancel the charged Higgs loop. In supersymmetric models, the mass of the charged Higgs, CP-odd pseudoscalar, and one CP-even scalar are set by a common mass scale with relatively small splittings. Thus, the high mass of the charged Higgs requires the Higgs at 125~GeV to be the lightest in the MSSM. 

Given the large number of possible parameters and the need for large loop corrections to obtain a 125~GeV aligned Higgs (as in the MSSM), we do not perform scans to fit to data, but restrict ourselves to a general discussion of the possibilities. In the large $\tan\beta$ limit of the four-Higgs supersymmetric model, the mass matrices for the Higgs sector become nearly block diagonal (barring very large $\mu_3$ and $\mu_4$ terms). As a result, the CP-even scalar with Standard Model-like couplings to the $W$ and $Z$ bosons (which we must identify with the 125~GeV Higgs discovered at the LHC) is primarily composed of an admixture of $H_u$ and $H_d$, as are the pseudoscalar and charged fields eaten by the $Z$ and $W$ bosons. The remaining $H_u$ and $H_d$ states, consisting of a CP-even neutral scalar, a CP-odd neutral pseudoscalar, and a charged scalar, must be heavy enough to avoid large flavor-changing decays in the quark sector, as in the MSSM. The leptophilic components, two each of the neutral scalars, pseudoscalars and charged pairs, can be a generic admixture of $H_\ell$ and $H_n$ fields, though if $\tan\gamma_\ell \gg 1$ or $\ll 1$, this will tend to result in each mass eigenstate being a nearly pure $H_\ell$ or $H_n$ state.  At tree level, two triads composed of a scalar, pseudoscalar, and charged Higgs will cluster in mass, with the scalar and pseudoscalar masses very close together, and the charged Higgs heavier by a $m_W^2$ addition to its mass squared.

These leptophilic Higgses can themselves be very light. The charged Higgses composed of $H_n^\pm$ and $H_\ell^\pm$ must be heavier than the LEP-II limit of $\sim 92$~GeV, but as seen in the L2HDM have very few other meaningful constraints. The LEP limit can be easily satisfied, even if the neutral components of the triad is well below $m_Z$. 

The CP-even and -odd components of a triad can be produced through their coupling to the $Z$, as in the L2HDM. Therefore, the LEP-II limits on $e^-e^+ \to Z^* \to H a \to 4\tau$, as discussed in Section~\ref{sec:overview}D apply here. As before, the sum of the scalar and pseudoscalar masses must be $\gtrsim 180$~GeV. Barring large loop corrections splitting the scalar and pseudoscalar, this effectively places a lower limit on the mass of these new Higgs pairs of $\sim 90$~GeV. 

However, given that large loop corrections must be applied to the Standard Model-like Higgs to lift its mass to 125~GeV, it is not implausible that the loop corrections from the matter sector can introduce large additional quartics to the leptophilic sector as well, splitting the mass of the scalar and pseudoscalar components. It is interesting to note that, in the L2HDM with supersymmetric-like couplings, we found that the pseudoscalar could not be below half of the 125~GeV Higgs mass without introducing non-standard decays that are experimentally ruled out. This situation can be mitigated in the supersymmetric case if the lightest pseudoscalar is an approximately equal admixture of $a_n$ and $a_\ell$. In this case the coupling to the Standard Model-like Higgs is suppressed due to a cancellation between the two components with opposite hypercharge. This does allow a leptophilic pseudoscalar in the $\lesssim 60$~GeV mass range preferred by $(g-2)_\mu$ and suggested by the Galactic Center anomaly, though we again stress that this requires a large splitting in the scalar-pseudoscalar masses, introduced by loops. 

The results from the L2HDM with fermionic dark matter added by hand can be ported to the supersymmetric version, where the dark matter sector arises naturally. A primarily-bino dark matter particle with ${\cal O}(10\%)$ admixture of higgsino components can easily be obtained when $M_1 \ll \tilde{\mu}_i \ll M_2$. Large $\tan\beta$, combined with a light pseudoscalar from the $H_\ell$ and $H_n$ doublets will result in annihilation cross sections consistent with the Galactic Center anomalies or thermal annihilation, while the direct detection constraints are easily satisfied due to the decoupling of the quark-like Higgses. As in the L2HDM, the light scalars and pseudoscalar mediators between the dark matter sector and the visible sector have extremely weak constraints from collider searches; as a result, the most easily obtained signature of this leptophilic supersymmetric model would be through the dark sector.

\section{Conclusions \label{sec:conclusions}}

Post-Higgs discovery, one of the high priority tasks at the LHC is to determine the full structure of the Higgs sector which is responsible for electroweak symmetry breaking. While the Higgs discovered at 125~GeV appears very consistent with the predictions of a single Higgs doublet in the Standard Model, our measurements are still only accurate at the $\sim 10\%$ level for couplings to the electroweak gauge bosons, and significantly weaker when it comes to direct measurements of the couplings to fermions. Much work remains to be done.

Two-Higgs-doublet models are a simple and attractive extension to the Standard Model Higgs sector. However, much of the experimental and theoretical effort has been dedicated to a particular version of such models, the Type-II model as found in the MSSM. Strong indirect limits can be set on such models, due to the coupling of both Higgs doublets to the quark sector. In particular, the charged Higgs must be heavy, which in the MSSM indicates that the discovered Higgs is the lightest of the physical scalars and the remaining Higgs are significantly decoupled.

In the light of the LHC results, we reconsider an alternative two-Higgs model, in which one doublet couples to both up- and down-type quarks while the other couples to the leptons only. This model easily evades most experimental searches for additional components of the Higgs sector, especially in the large $\tan\beta$ regime where the new mass eigenstates  couple mostly to leptons, rather than quarks. Additional Higgs scalars can be lighter than the 125~GeV CP-even Higgs, with the strongest constraints coming from LEP-II. These limit the charged Higgs to be heavier than $\sim 92$~GeV, and require the sum of the additional CP-even and CP-odd Higgs masses to be greater than $180-200$~GeV. 

Given the relative alignment of the 125~GeV Higgs, should leptophilic Higgses exist, their discovery at the LHC will be  difficult. From the LHC Higgs measurements, we now know that the most promising channels, considered previously \cite{Marshall:2010qi, Logan:2009uf} would have small cross sections. Direct production of the scalar and pseudoscalar particles through VBF or gluon fusion process is suppressed by $\cos(\beta-\alpha)^2$ or $\cot^2\beta$. The pair production mechanism of a scalar and a pseudoscalar through the $Z$ followed by decays to tau lepton pairs is not suppressed, but is an extremely difficult search given the efficiency for tau-tagging and the large backgrounds.  For example, we have demonstrated that the existing multilepton analysis from CMS in Run-I \cite{Khachatryan:2014jya} is not sensitive to scalar/pseudoscalar pairs with masses at the LEP-II bound. The increased cross section at Run-II will be helpful, but a full analysis of the tau backgrounds will be necessary and the channel will remain difficult. New ideas may be necessary.

Indeed, the discovery of leptophilic Higgses might be most easily achieved through its coupling to dark matter. Motivated by supersymmetric extensions of the leptophilic model, we consider fermionic dark matter composed of $SU(2)_L$ doublets mixed with singlets -- analogs of higgsino-bino dark matter.   In the large $\tan\beta$ limit, the dark matter has significant interactions with the leptophilic Higgs scalars, through the doublet component. We find significant parameter space for thermal dark matter, with possible masses extending below half the mass of the 125~GeV Higgs or the $Z$ mass without violating the invisible width constraints on those particles. Dark matter indirect detection rates are naturally suppressed compared to the expectations from the Higgs portal in Type-II model. However, barring accidental cancellations, the direct detection rate is in reach of the next generation of direct detection experiments. The large couplings to the leptophilic pseudoscalar allow for significant $s$-wave annihilation in the Universe today, allowing for a 10~GeV dark matter particle annihilating into tau leptons, as has been suggested as a possible fit to the Galactic Center anomaly with mediator masses $\lesssim 90$~GeV.

We revisit the supersymmetric extension of this model, including four Higgs doublets, which is leptophilic in the limit where the majority of the EWSB vev resided in the quark sector. As in the MSSM, significant loop-corrections are required in order to bring the Standard Model-like Higgs up to the observed mass. Thus, though the supersymmetric leptophilic model allows the interesting possibility of new Higgses lighter than the one already discovered, it does not address the Little Hierarchy Problem. The attractive properties of dark matter in a leptophilic two-Higgs model are found as well in the supersymmetric version: low mass mediators capable of fitting the Galactic Center anomaly would not have been seen in colliders, and unlike in the MSSM, would not induce large direct detection signals in tension with experimental results. Mixing between the two leptophilic Higgs doublets can also allow a pseudoscalar lighter than half the 125~GeV Higgs mass, which is not possible in the leptophilic two-Higgs model. As in the L2HDM, the discovery of the extended Higgs sector by itself is difficult at the LHC, as the production cross sections most conducive to LHC searches are suppressed in a leptophilic Higgs model with one nearly-Standard Model-like Higgs, as is the experimental situation we find ourselves. The addition of supersymmetric particles allows for larger production cross sections, as colored particles will undergo cascade decays into the neutralino/chargino sector. These cascades would preferentially  include tau leptons in the supersymmetric leptophilic model. Though searches including tau final states have been performed, they are experimentally challenging, and existing constraints are weak and the potential for improvement is large.

\end{document}